# An axis symmetric 2D description of the process of the growth of a single crystal Si tube growth from the melt by pulling down method.Part1.

Agneta M. Balint[1], Stefan Balint[2] , Loredana Tanasie[2]
[1]*Department of Physics, West University of Timisoara, Bulv. V. Parvan 4, 300223 Timisoara, Romania*
[2] *Department of Computer Science, West University of Timisoara, Bulv. V. Parvan 4, 300223 Timisoara, Romania*
*Corresponding author: Loredana Tanasie*
*Department of Computer Science, West University of Timisoara, 300223 Timisoara, Bulv. V. Parvan 4, Romania*
*stefan.balint@e-uvt.ro*

**Abstract.** This paper is the first part of an axis symmetric 2D description of a single crystal Si tube growth process by micro- pulling–down (PD) method. The description concerns the following aspects: the free surfaces equations and the pressure differences across the free surfaces (section 2.); limits of the pressure differences $p_i$ and $p_e$ across the free surfaces (section 3.); static stability of the capillary free surfaces (section 4). In section 5 the above aspects are numerically investigated in the case of the growth of a Si micro tube of inner radius equal to $4.28x10^{-3}$[m] and outer radius equal to $4.72x10^{-3}$ by using the Maple 17 software. This investigation can be helpful in the better understanding of the growth process and in the automation of manufacturing. In section 6 some general conclusion are given. Section 7 is an annex which contains the proofs of the general statements formulated in sections 3 and 4.

## 1. INTRODUCTION

The micro-pulling-down technique is a crystal growth method that has been mostly developed since 1992. [1], [2]. The general scheme of the growth system is the following: the melt (oxide, fluoride, metal) residing in a crucible is transported in downward through a capillary channel placed in the bottom of the crucible. Two driving forces (capillary action and gravity) support the delivery of the melt to the shaper placed at the capillary channel bottom. Appropriate configuration of the shaper (for tube annulus) allows for controlling of the crystal shape (fibers, rods, tubes, plates) and the dimensions of the crystals cross sections that range approximately from 0.1 to 10 mm. By introducing a seed, of the crystal being to grown, in the bottom of the shaper melt, a small part of the seed will be melted and a liquid bridge appears between the shaper melt and the non melted part of the seed. This liquid bridge is called meniscus. By pulling down slowly the seed and maintaining some appropriate thermal conditions (cooling the meniscus and the seed) the melt begins to flow down and solidifies . A great number of scientifically and industrially important optical crystal fibers, rods, tubes, plates have been successfully produced using this method. The papers[1]-[6] represent just a small part of the results reported in the scientific literature.

Fig.1 is a schematic representation of the cylindrical tube growth by micro-pulling-down method.

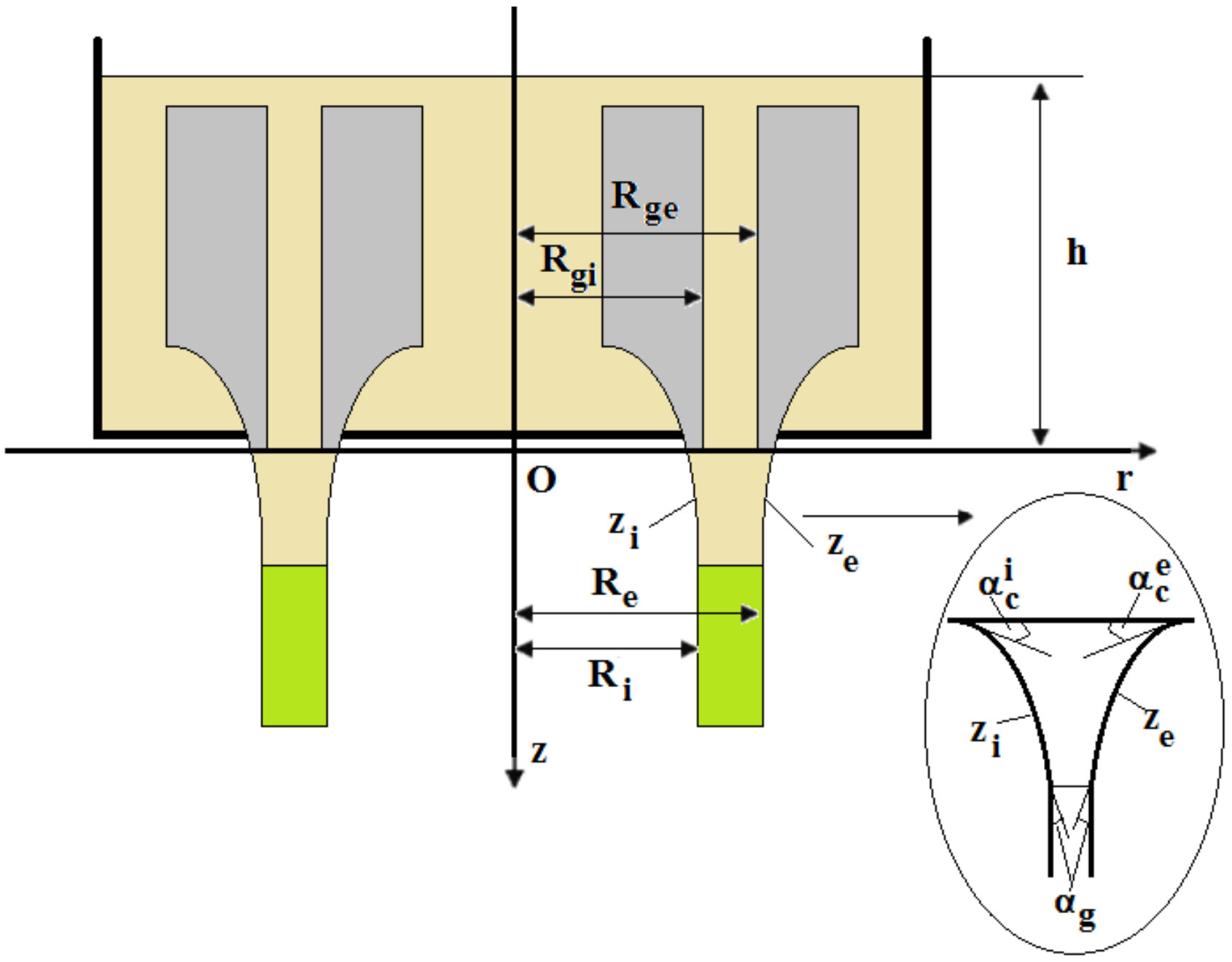

Fig.1. Schematic representation of the cylindrical tube growth by micro-pulling-down method.

In the present paper the following aspects related to single crystal cylindrical tube growth from the melt by PD method in a 2D axis-symmetric configuration (Fig.1) are presented: the capillary free surfaces equations and the pressure differences across the free surfaces (section2), limits of the pressure differences across the capillary free surfaces (section 3), static stability/instability of the capillary free surfaces (section4), numerical investigations in case of a Si tube growth by PD method (section 5), conclusions (section 6), proof of some theoretical results (Annex).

## 2.THE CAPILLARY FREE SURFACES EQUATIONS AND THE PRESSURE DIFFERENCES ACROSS THE FREE SURFACES IN A P.D. GROWTH PROCESS OF A TUBE.

In case of tube growth from the melt by PD method (see Fig.1) the meniscus has two capillary free surfaces. One of them is in the interior of the tube the other is in the exterior of the tube. These free surfaces are called inner and outer free surfaces. Both of them are described mathematically by the Young-Laplace equation [7]-Chapter1,[8] –Chapter 1:

$$\gamma \cdot (1/R_1 + 1/R_2) = P_a - P_b \qquad (2.1)$$

Here: $\gamma$ is the melt surface tension; $1/R_1$, $1/R_2$ denote the main normal curvatures of the inner (outer) free surface at a point M of the free surface; $P_a$ is the pressure in front of the inner (outer) free surface; $P_b$ is the pressure behind the inner (outer) free surface (Fig.1). Thereafter the pressure $P_a$ for the inner free surface is denoted by $P_a = P_a{}^i$ and for the outer free surface is denoted by $P_a = P_a{}^e$ .

The pressure $P_b$ behind the inner /outer free surface is the sum of the hydrodynamic pressure $p_m$ in the meniscus melt (due to the convection), the Marangoni pressure $p_M$ due to the thermal Marangoni convection and the hydrostatic pressure of the melt column behind the free surface equal to $\rho \cdot g \cdot (z + h_m)$ (see Fig.1). Here: $\rho$ denotes the melt density; $g$ is the gravity acceleration; $z$ is the coordinate of M with respect to the $Oz$ axis, directed vertically downwards; $h_m$ denotes the melt column height between the horizontal crucible melt level and the shaper bottom level (Fig.1).

For the inner free surface $P_b = P_b{}^i = p^i{}_m + p^i{}_M + \rho \cdot g \cdot (z^i + h^i{}_m)$ and for the outer free surface $P_b = P_b{}^e = p^e{}_m + p^e{}_M + \rho \cdot g \cdot (z^e + h^e{}_m)$

The pressure difference $P_a - P_b$ across the inner /outer free surface is denoted usually by $\Delta p$ and, according to the above considerations

$$\Delta p = P_a - p_m - p_M - \rho \cdot g \cdot (z + h_m)\,. \qquad (2.2)$$

For the inner free surface $\Delta p^i = P_a{}^i - p^i{}_m - p^i{}_M - \rho \cdot g \cdot (z^i + h^i{}_m)$ and for the outer free surface $\Delta p^e = P_a{}^e - p^e{}_m - p^e{}_M - \rho \cdot g \cdot (z^e + h^e{}_m)$.

The part $p$ of the inner /outer total pressure difference $-\Delta p$ given by:

$$p = -P_a + \rho \cdot g \cdot h_m + p_m + p_M \qquad (2.3)$$

is independent of the spatial coordinate $z$ (it can depend on the moment of time $t$) and it is assumed that the major part of $p$ is $-P_a + \rho \cdot g \cdot h_m$. That is because it is assumed that the sum $p_m + p_M$ in general is small with respect to $-P_a + \rho \cdot g \cdot h_m$. For this reason thereafter it is assumed that the inner /outer pressure difference $p$ is given by:

$$p = -P_a + \rho \cdot g \cdot h_m \qquad (2.4)$$

For the inner free surface $p_i = -P^i{}_a + \rho \cdot g \cdot h^i{}_m$ and for the outer free surface $p_e = -P^e{}_a + \rho \cdot g \cdot h^e{}_m$.

Remark that in case of the inner free surface the pressure difference $p_i$ can be controlled by $P_a{}^i$ and $h^i{}_m$. In case of the outer free surface the pressure difference $p_e$ can be controlled by $P_a{}^e$ and $h^e{}_m$.

With this approximation the Young-Laplace equation (2.1) for the inner/ outer free surfaces can be written as:

$$\gamma \cdot \left(1/R^i{}_1 + 1/R^i{}_2\right) = -\rho \cdot g \cdot z_i - p_i \qquad \gamma \cdot \left(1/R_1^e + 1/R^e{}_2\right) = -\rho \cdot g \cdot z_e - p_e \qquad (2.5)$$

respectively.

Therefore, for an axis-symmetric 2D meniscus free surface the differential equation of the meridian curves of the inner and outer free surfaces with respect to the coordinate frame indicated in Fig.1 are given by:

$$z_i'' = -\frac{\rho \cdot g \cdot z_i + p_i}{\gamma}\left[1 + (z_i')^2\right]^{3/2} - \frac{1}{r}\cdot\left[1 + (z_i')^2\right]\cdot z_i'\,; \quad R_{gi} \le r \le R_{ci} \qquad (2.6)$$

$$z_e'' = -\frac{\rho \cdot g \cdot z_e + p_e}{\gamma}\left[1 + (z_e')^2\right]^{3/2} - \frac{1}{r}\cdot\left[1 + (z_e')^2\right]\cdot z_e'\,; \quad R_{ce} \le r \le R_{ge} \qquad (2.7)$$

Here: $R_{gi}, R_{ge}$ represent the inner and outer radius of the shaper ; $R_{ci}$ , $R_{ce}$ represent the inner and outer radius of the tube being to grown and satisfies $0 < R_{gi} < R_{ci} < \frac{R_{gi} + R_{ge}}{2}$, $\frac{R_{gi} + R_{ge}}{2} < R_{ce} < R_{ge}$ , respectively.

The function $z_i(r)$, describing the inner free surface meridian curve profile, verifies the following boundary conditions:

$$z_i'(R_{ci}) = \tan(\pi/2 - \alpha_g)\,; \qquad z_i(R_{ci}) = h^i{}_c > 0 \qquad (2.8)$$

$$z_i'(R_{gi}) = \tan \alpha_c\,; \qquad z(R_{gi}) = 0 \qquad (2.9)$$

$z_i(r)$ is strictly increasing on $[R_{gi}, R_{ci}]$ (2.10)

The first condition in (2.8) expresses that at the three phase point $(R_{ci}, h^i{}_c)$, where the thermal conditions for solidification have to be realized ($h^i{}_c$ is the inner crystallization front level), the angle between the tangent line to the meridian curve of the free surface and the vertical is equal to the growth angle $\alpha_g$. If this condition is satisfied during the whole growth process, then the tube inner diameter is constant. The second condition in (2.8) expresses that the coordinate of the crystallization front level is $h^i{}_c > 0$.

The first condition in (2.9) expresses that at the point $(R_{gi}, 0)$, where the meridian curve touches the outer edge of the shaper, the catching angle (contact angle) is equal to $\alpha^i{}_c$. The second condition in (2.9) expresses that at the point $(R_{gi}, 0)$ the meridian curve is fixed to the outer edge of the shaper.

Condition (2.10) expresses that the meniscus shape is relatively simple.

Beside the conditions (2.6), (2.8), (2.9) and (2.10) the function $z_i(r)$ describing the inner meridian curve profile has to minimize the free energy functional, of the melt column behind the inner capillary free surface, given by :

$$I(z_i) = \int_{R_{gi}}^{R_{ci}} \left\{ \gamma \cdot \left[1 + (z_i')^2\right]^{1/2} - \frac{1}{2} \cdot \rho \cdot g \cdot z_i^2 - p_i \cdot z_i \right\} \cdot r \cdot dr \qquad (2.11)$$

This condition is the static stability of the axis symmetric inner capillary free surface. Note that in real world only static stable capillary surfaces exist.

The function $z_e(r)$, describing the outer free surface meridian curve profile, verifies (2.7) and the following boundary conditions:

$$z_e'(R_{ce}) = -\tan(\pi/2 - \alpha_g); \qquad z_e(R_{ce}) = h^e{}_c > 0 \qquad (2.12)$$

$$z_e'(R_{ge}) = -\tan \alpha_c; \qquad z_e(R_{ge}) = 0 \qquad (2.13)$$

$$z_e(r) \text{ is strictly increasing on } [R_{ce}, R_{ge}] \qquad (2.14)$$

The first condition in (2.12) expresses that at the three phase point $(R_{ce}, h^e{}_c)$, where the thermal conditions for solidification have to be realized ($h^e{}_c$ is the outer crystallization front level), the angle between the tangent line to the meridian curve of the free surface and the vertical is equal to the growth angle $\alpha_g$. If this condition is satisfied during the whole growth process, then the tube outer diameter is constant. The second condition in (2.12) expresses that the coordinate of the outer crystallization front level is $h^e{}_c > 0$.

The first condition in (2.13) expresses that at the point $(R_{ge}, 0)$, where the meridian curve touches the outer edge of the shaper, the catching angle (contact angle) is equal to $\alpha_c$. The second condition in (2.13) expresses that at the point $(R_{ge}, 0)$ the meridian curve is fixed to the outer edge of the shaper.

Condition (2.14) expresses that the meniscus shape is relatively simple.

Beside the Eq.2.7 and conditions (2.12), (2.13) and (2.14), the function $z_e(r)$ describing the outer meridian curve profile of the outer capillary free surface has to minimize the free energy functional, of the melt column behind the outer capillary free surface, given by :

$$I(z_e) = \int_{Rce}^{R_{ge}} \left\{ \gamma \cdot \left[1 + (z_e')^2\right]^{1/2} - \frac{1}{2} \cdot \rho \cdot g \cdot z_e^2 - p_e \cdot z_e \right\} \cdot r \cdot dr \qquad (2.15)$$

This condition assure the static stability of the axis symmetric outer capillary free surface. Note that in real world only static stable capillary free surfaces exist.

## 3. LIMITS OF THE PRESSURE DIFFERENCES $p_i$ AND $p_e$ IN A PD GROWTH PROCESS

If $p_i$ $p_e$ are arbitrary, then it can happen that there are no functions $z_i(r), z_e(r)$ which verify (2.6) - (2.15) .This means that there is no convex static axis-symmetric 2D meniscus. The following statement is a necessary condition for the existence of a function $z_i(r)$ having the properties (2.6), (2.8)- (2.10) and $z_i''(r) > 0$ for $r \in \lfloor R_{gi}, R_{ci} \rfloor$.

**Statement 1.** If $\alpha_c + \alpha_g < \pi/2$, $R_{ci} = m \cdot R_{gi}$, $1 < m < \dfrac{R_{gi} + R_{ge}}{2 \cdot R_{gi}}$, then in order to have an axis-symmetric 2D meniscus whose inner free surface has convex meridian curve ($z_i''(r) > 0$) on the interval $\lfloor R_{ci} = m \cdot R_{gi}, R_{gi} \rfloor$ the pressure $p_i$ has to satisfy the following inequalities:

$$\gamma \cdot \frac{\alpha_c + \alpha_g - \pi/2}{R_{gi}} \cdot \frac{1}{m-1} \cdot \cos\alpha_c - \rho \cdot g \cdot R_{gi} \cdot (m-1) \cdot \tan(\pi/2 - \alpha_g) - \frac{\gamma}{R_{gi}} \cdot \sin\alpha_g \leq p_i \leq \gamma \cdot \frac{\alpha_c + \alpha_g - \pi/2}{R_{gi}} \cdot \frac{1}{m-1} \cdot \sin\alpha_g - \frac{\gamma}{R_{gi}} \cdot \frac{1}{m} \cdot \sin\alpha_c \quad (3.1)$$

The proof of the above statement can be found in the Annex.

**Comment 1:** According to **Statement 1**, in order to obtain an axis-symmetric 2D meniscus whose inner free surface has convex meridian curve on the interval $\lfloor R_{ci} = m \cdot R_{gi}, R_{gi} \rfloor$ the range where the pressure difference $p_i$ has to be taken , is the pressure interval $[L_{1i}(m),\ L_{2i}(m)]$ [Pa] where:

$$L_{1i}(m) = \gamma \cdot \frac{\alpha_c + \alpha_g - \pi/2}{R_{gi}} \cdot \frac{1}{m-1} \cdot \cos\alpha_c - \rho \cdot g \cdot R_{gi} \cdot (m-1) \cdot \tan(\pi/2 - \alpha_g) - \frac{\gamma}{R_{gi}} \cdot \sin\alpha_g \quad (3.2)$$

$$L_{2i}(m) = \gamma \cdot \frac{\alpha_c + \alpha_g - \pi/2}{R_{gi}} \cdot \frac{1}{m-1} \cdot \sin\alpha_g - \frac{\gamma}{R_{gi}} \cdot \frac{1}{m} \cdot \sin\alpha_c \quad (3.3)$$

There is no reason to search the pressure difference $p_i$ outside of the above interval.

The range $[L_{1i}(m),\ L_{2i}(m)]$ where the value of pressure difference $p_i$ has to be, can be large and we don't know which value of $p_i$ from this range is appropriate. Moreover, is not sure that there exists a value of $p_i$ in the above range for which an inner convex meridian curve exists (the condition (3.1) is only necessary).

In order to answer these questions the following initial value problem:

$$\begin{cases} \dfrac{dz_i}{dr} = \tan\alpha_i \\ \dfrac{d\alpha_i}{dr} = -\dfrac{\rho \cdot g \cdot z_i + p_i}{\gamma} \cdot \dfrac{1}{\cos\alpha_i} - \dfrac{1}{r} \cdot \tan\alpha_i \\ z(R_{gi}) = 0, \quad \alpha(R_{gi}) = \alpha_c \end{cases} \quad (3.4)$$

has to be solved numerically for different values of the pressure difference $p_i$ in the range $[L_{1i}(m), L_{2i}(m)]$[Pa]. In this way can be found (if exists) that value of $p_i$ for which $R_{ci} = m \cdot R_{gi}[m]$ and $\alpha_i(R_{ci}) = \pi/2 - \alpha_g$ $[rad]$.

The next statement is a necessary condition for the existence of a function $z_e(r)$ having the properties (2.7), (2.12) - (2.14) and $z_e''(r) > 0$ for $r \in \lfloor R_{ce}, R_{ge} \rfloor$.

**Statement 2.** If $\alpha_c + \alpha_g < \pi/2$, $R_{ce} = \frac{R_{ge}}{n}$, $1 < n < \frac{2 \cdot R_{ge}}{R_{gi} + R_{ge}}$, then in order to have an axis-symmetric 2D meniscus whose outer free surface has convex meridian curve ($z_e''(r) > 0$) on the interval $\left[ R_{ce} = \frac{R_{ge}}{n}, R_{ge} \right]$ the pressure $p_e$ has to be chosen in the range

$$\gamma \cdot \frac{\alpha_c + \alpha_g - \pi/2}{R_{ge}} \cdot \frac{n}{n-1} \cdot \cos\alpha_c - \rho \cdot g \cdot R_{ge} \cdot \frac{n-1}{n} \cdot \tan(\pi/2 - \alpha_g) - \frac{\gamma}{R_{ge}} \cdot n \cdot \cos\alpha_g \le p_e \le \gamma \cdot \frac{\alpha_c + \alpha_g - \pi/2}{R_{ge}} \cdot \frac{n}{n-1} \cdot \sin\alpha_g - \frac{\gamma}{R_{ge}} \cdot \sin\alpha_c \tag{3.5}$$

The proof of the above statement can be proven in the Annex.1.

**Comment 2:** According to **Statement 2**, in order to obtain an axis-symmetric meniscus whose outer free surface has convex meridian curve on the interval . $\left[ R_{ce} = \frac{R_{ge}}{n}, R_{ge} \right]$ the range where the pressure difference $p_e$ has to be taken is the pressure interval $[L_{1e}(n),\ L_{2e}(n)]$ [Pa] where

$$L_{1e}(n) = \gamma \cdot \frac{\alpha_c + \alpha_g - \pi/2}{R_{ge}} \cdot \frac{n}{n-1} \cdot \cos\alpha_c - \rho \cdot g \cdot R_{ge} \cdot \frac{n-1}{n} \cdot \tan(\pi/2 - \alpha_g) - \frac{\gamma}{R_{ge}} \cdot n \cdot \cos\alpha_g \tag{3.6}$$

$$L_{2e}(n) = \gamma \cdot \frac{\alpha_c + \alpha_g - \pi/2}{R_{ge}} \cdot \frac{n}{n-1} \cdot \sin\alpha_g - \frac{\gamma}{R_{ge}} \cdot \sin\alpha_c \tag{3.7}$$

There is no reason to search the pressure difference $p_e$ outside of the above interval.

The range $[L_{1e}(n),\ L_{2e}(n)]$[Pa] where the values of the pressure difference $p_e$ has to be, can be large and we don't know which value of $p_e$ from this range is appropriate. Moreover, is not sure that there exists a value of $p_e$ in the above range for which an inner convex meridian curve exists (the condition (3.5) is only necessary).

In order to answer these questions the following initial value problem:

$$\begin{cases} \frac{dz_e}{dr} = -\tan\alpha_e \\ \frac{d\alpha_e}{dr} = \frac{\rho \cdot g \cdot z_e + p_e}{\gamma} \cdot \frac{1}{\cos\alpha_e} + \frac{1}{r} \cdot \tan\alpha_e \\ z(R_{ge}) = 0, \quad \alpha(R_{ge}) = \alpha_c \end{cases} \tag{3.8}$$

has to be solved numerically for different values of the pressure difference $p_e$ in the range of the pressure difference $[L_{1e}(n), L_{2e}(n)]$ [Pa]. In this way can be found (if exist) that value of $p_e$ for which $R_{ce} = \frac{R_{ge}}{n}[m]$ and $\alpha_e(R_{ce}) = \pi/2 - \alpha_g\ [rad]$.

## 4. STATIC STABILITY OF THE FREE SURFACE

This section deals with the static stability/instability of the capillary free surfaces appearing in tube growth from the melt by PD method. The static stability of the capillary free surfaces is an absolutely necessary condition because in real world only static stable capillary free surfaces exist.
The following statement is a sufficient condition of static stability/instability of the inner capillary free surface.

**Statement 3.** If $\alpha_c + \alpha_g < \pi/2$, $R_{ci} = m \cdot R_{gi}$, $1 < m < \frac{R_{gi} + R_{ge}}{2 \cdot R_{gi}}$ and there exits $p_i$ in the range of the pressure difference $[L_{1i}(m), L_{2i}(m)]$ [Pa] for which the solution $z_i(r)$ of (2.6) and (2.8)-(2.10) satisfies $z_i''(r) > 0$ on the interval $\lfloor R_{gi}, m \cdot R_{gi} \rfloor$, then:
for

$$\frac{m-1}{m^{1/2}} < \pi \cdot \frac{1}{R_{gi}} \cdot \frac{\gamma^{1/2} \cdot \sin^{3/2} \alpha_g}{\rho^{1/2} \cdot g^{1/2}} \tag{4.1}$$

$z_i(r)$ minimizes the free energy functional $I_i(z)$ of the melt column behind the inner capillary free surface and the free surface is static stable and for

$$\frac{m-1}{m^{3/2}} > 2 \cdot \pi \cdot \frac{1}{R_{gi}} \cdot \frac{\gamma^{1/2} \cos^{3/2} \alpha_c}{\rho^{1/2} \cdot g^{1/2}} \tag{4.2}$$

the inner free surface is static unstable.
**Comment 3:** If $R_{gi}$ is decreased, then the right hand member of (4.1) increases and so stability can appear, but if $R_{gi}$ is increased, then the right hand member of (4.2) decreases and so instability can appear.
For materials with small growth angle stability appears for small $R_{gi}$, but for materials with big contact growth angle instability appears for big $R_{gi}$

The following statement is a sufficient condition of static stability/instability of the outer capillary free surface.

**Statement 4.** If $\alpha_c + \alpha_g < \pi/2$, $R_{ce} = \frac{R_{ge}}{n}$, $1 < n < \frac{2 \cdot R_{ge}}{R_{gi} + R_{ge}}$ and there exits $p_e$ in the range of the pressure difference $[L_{1e}(n), L_{2e}(n)]$ [Pa] for which the solution $z_e(r)$ of (2.7) and (2.12)-(2.14) satisfies $z_e''(r) > 0$ on the interval $\left[ R_{ce} = \frac{R_{ge}}{n}, R_{ge} \right]$, then:
for

$$\frac{n-1}{n^{1/2}} < \pi \cdot \frac{1}{R_{ge}} \cdot \frac{\gamma^{1/2} \cdot \sin^{3/2} \alpha_g}{\rho^{1/2} \cdot g^{1/2}} \ , \tag{4.3}$$

$z_e(r)$ minimizes the free energy functional $I_e(z)$ of the melt column behind the outer capillary free surface and the outer free surface is static stable
for

$$\frac{n-1}{n^{3/2}} > 2\pi \cdot \frac{1}{R_{ge}} \cdot \frac{\gamma^{1/2} \cdot \cos^{3/2} \alpha_c}{\rho^{1/2} \cdot g^{1/2}} \ , \tag{4.4}$$

the outer free surface is static unstable.
**Comment 4**. If $R_{ge}$ is decreased, then the right hand member of (4.3) increases and so stability appears but if $R_{ge}$ is increased, then the right hand member of (4.4) decreases and so instability appears.

For materials with small growth angle stability appears for small $R_{gi}$ , but for materials with big contact growth angle instability appears for big $R_{gi}$

## 5. NUMERICAL INVESTIGATION.

In this section the presented theoretical results will be numerically investigated in case of a Si tube grown from the melt by PD. The numerical data are :

$$R_{gi} = 4.2\times10^{-3}[m];\ R_{ge} = 4.8\times10^{-3}[m]\ \ \alpha_c = 0.523\ [rad];\ \alpha_g = 0.192\ [rad];\ \rho = 2.58\times10^{3}\ [kg/m^{3}];$$

$$g = 9.81\ [m/s^{2}]\ ;\ \gamma = 0.765[N/m]\ ;R_{ci} = 4.28\times10^{-3}[m]\ \text{and}\ R_{ce} = 4.72\times10^{-3}[m].$$

**i). Computation of the pressure difference limits**

For the considered data the computed values of $m$ and $n$ are $m = \frac{R_{ci}}{R_{gi}} = 1.0190$ and $n = \frac{R_{ge}}{R_{ce}} = 1.0164$. Replacing $m$ in formulas (3.2) and (3.3) the following limits for the inner pressure difference $p_i$ across the inner capillary surface are obtained:

$$-6866.968527\,[Pa] \le p_i \le -1557.459343\,[Pa]$$

Replacing $n$ in formulas (3.6) and (3.7) the following limits for the outer pressure difference $p_e$ across the outer capillary surface are obtained

$$-7035.557193\,[Pa] \le p_e \le -1589.429076\,[Pa]$$

**ii). Computation of the exact values of the pressure differences $p_i$ and $p_e$**

By solving numerically the initial value problem (3.4), for different values of the pressure difference $p_i$ situated in the range $[-6866.968527, -1557.459343][\mathrm{Pa}]$, is found that the value of $p_i$ for which $R_{ci} = 4.28\times10^{-3}[m]$ and $\alpha_i(R_{ci}) = \pi/2 - \alpha_g\ [rad] = 1.378[rad]$ is $p_i = -4745[Pa]$. The computed inner meridian curve height for $p_i = -4745[Pa]$ is $h^{i}_{c} = z_i(\mathrm{R_{ci}}) = 1.124052\cdot10^{-4}\,[m]$.The meridian curve is presented in Fig.2.a and the variation of $\alpha_i(r)$ in Fig.2b.

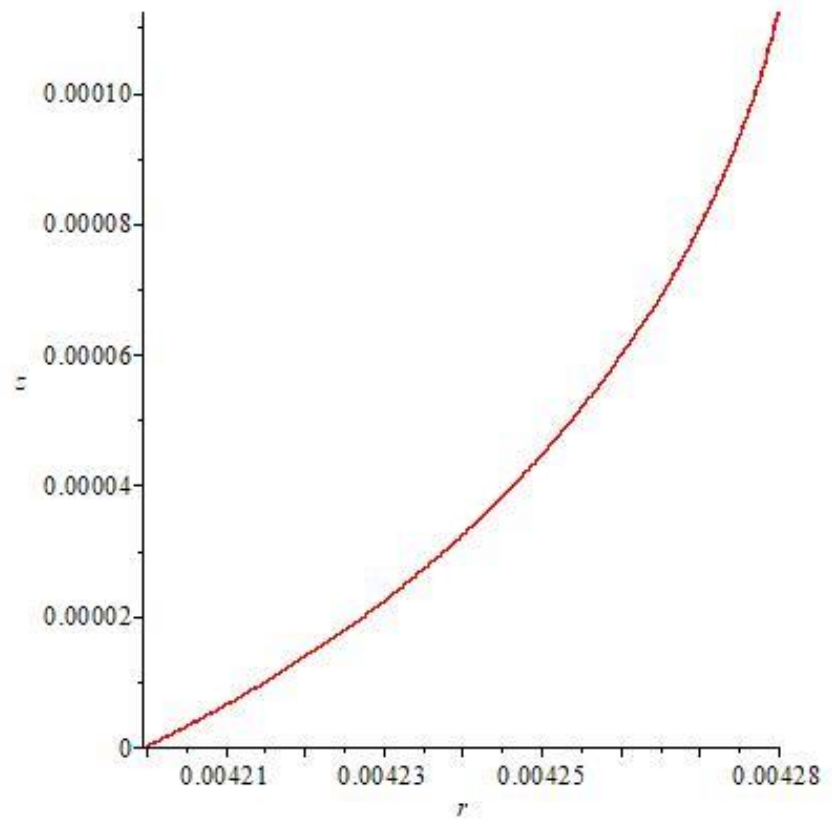

Fig.2a. Shape of $z_i(r)$ for $p_i = -4745[Pa]$

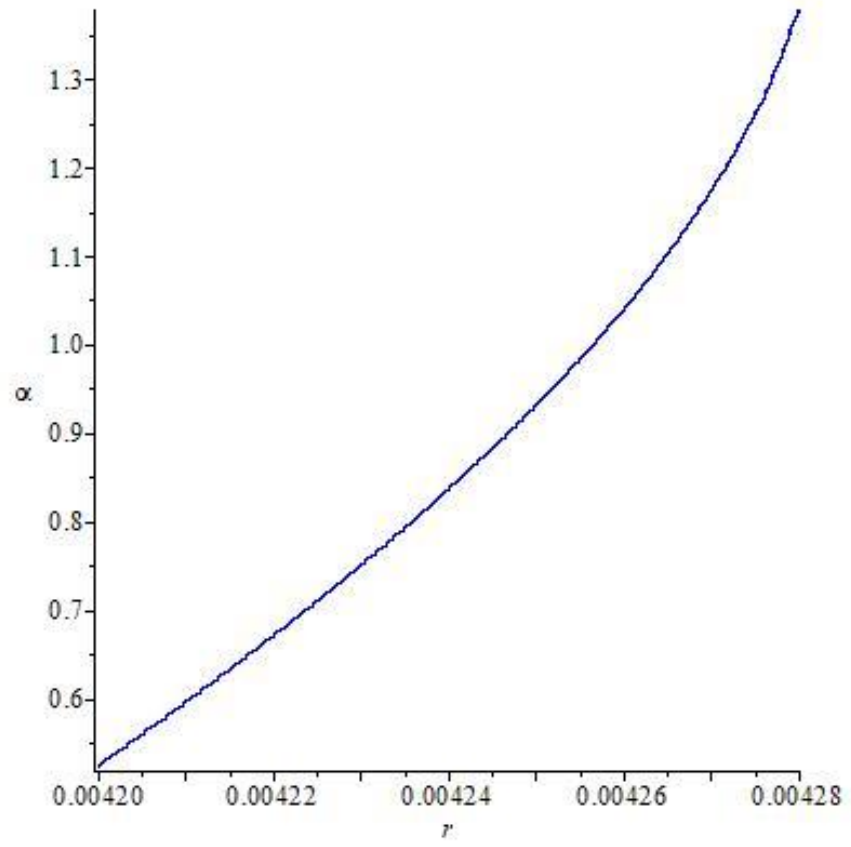

Fig.2b. Shape of $\alpha_i(r)$ for $p_i = -4745[Pa]$

By solving numerically the initial value problem (3.8), for different values of the pressure difference $p_e$ situated in the range $[-7035.557193, -1589.429076][\mathrm{Pa}]$, is found that the value of $p_e$ for which $R_{ce} = 4.76\times10^{-3}[m]$ and $\alpha_e(R_{ce}) = \pi/2 - \alpha_g\ [rad] = 1.378[rad]$ is $p_e = -4730[Pa]$. The computed outer meridian curve height for $p_e = -4730[Pa]$ is $h^e{}_c = z_e(R_{ce}) = 1.123473 \cdot 10^{-4}\,[m]$. The meridian curve $z_e(r)$ is presented in Fig.3.a and the variation of $\alpha_e(r)$ in Fig.3b.

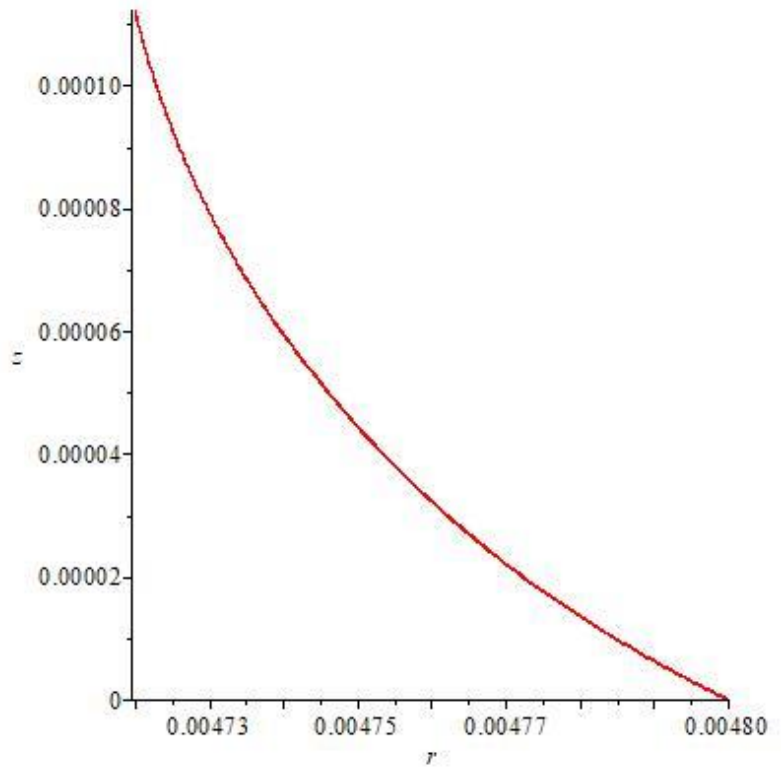

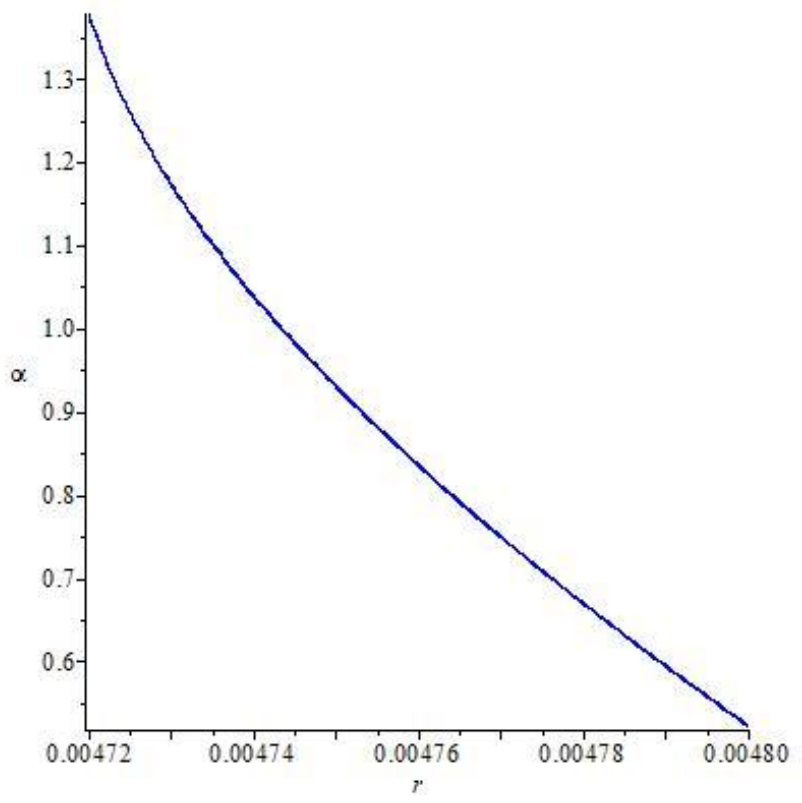

Fig.3a. Shape of $z_e(r)$ for $p_e = -4730\,[Pa]$

Fig.3b. Shape of $\alpha_e(r)$ for $p_e = -4730\,[Pa]$

At this point it has to be noted that the downward orientation of the OZ axis in Fig.1. and the upward orientation of the OZ axis in figures Fig.2a; Fig.3a; are opposite. For that Fig2.a, Fig3.a have to be rotated with 180 degrees around the Or axis in order to obtain the meridian curve shape as is presented in Fig.1.

**iii). Static stability/instability of the inner and outer capillary surfaces**

Inequality (4.1) in this case becomes 0.01882203206< 0.3325576396, showing that the inner capillary free surface is static stable.

Inequality (4.3) in this case becomes 0.01626715182< 0.2909879348 showing that the outer capillary free surface is static stable.

**iv). The effect of the variation of pressure difference** $p_i$

Starting from the condition of growth angle constancy, according to [26]-Chapter 1, the equation of the inner radius change rate, due to the perturbation of pressure difference $-4745[Pa]$, is:

$$\begin{cases} dR_{ci}/dt = -v\tan(\alpha_i(4.28 \cdot 10^{-3}, p_i) - (\pi/2 - \alpha_g)) \\ R_{ci}(0) = 4.28 \cdot 10^{-3} \end{cases} \tag{5.1}$$

Here $R_{ci}(t)$ is the inner radius at the moment of time $t$, $v = 2 \cdot 10^{-3}[m/s]$ is the pulling rate used in Eris experiment[1] , $\alpha_i(4.28 \cdot 10^{-3}, p_i)$ is the angle made by the line tangent to the perturbed inner meridian curve at the point of coordinates $(4.28 \cdot 10^{-3}, z_i(4.28 \cdot 10^{-3}, p_i))$ ,with the horizontal axis and $p_i$ is the perturbed inner pressure difference.

If $p_i = -4745\,[Pa]$, then $\alpha_i(4.28 \cdot 10^{-3}, p_i) = \pi/2 - \alpha_g$.

So, $\tan(\alpha_i(4.28 \cdot 10^{-3}, p_i) - (\pi/2 - \alpha_g)) = \tan(0) = 0$ and (5.1) becomes:

$$\begin{cases} dR_{ci}/dt = 0 \\ R_{ci}(0) = 4.28 \cdot 10^{-3} \end{cases} \tag{5.2}$$

If $p_i = -4845\,[Pa]$, then $\alpha_i(4.2 \cdot 10^{-3}, p_i) = 1.44400[\text{ra d}]$.

So, $-v \cdot \tan(\alpha_i(4.28 \cdot 10^{-3}, p_i) - (\pi/2 - \alpha_g)) = -v \cdot \tan(0.06520)$ and (5.1) becomes:

$$\begin{cases} dR_{ci}/dt = -v\tan(0.06520) \\ R_{ci}(0) = 4.28 \cdot 10^{-3} \end{cases} \tag{5.3}$$

If $p_i = -4645\,[Pa]$, then $\alpha_i(4.28\cdot 10^{-3}, p_i) = 1.33045[rad]$.
So, $-v\cdot\tan(\alpha_i(4.28\cdot 10^{-3}, p_i) - (\pi/2 - \alpha_g)) = -v\cdot\tan(-0.04834)$ and (5.1) becomes:

$$\begin{cases} dR_{ci}/dt = v\cdot\tan(0.04834) \\ R_{ci}(0) = 4.2\cdot 10^{-3} \end{cases} \tag{5.4}$$

For $v = 2\cdot 10^{-3}[m/s]$ the solutions of the initial value problems (5.2),(5.3) and (5.4) are represented on Fig.4 with green, red and blue, respectively.

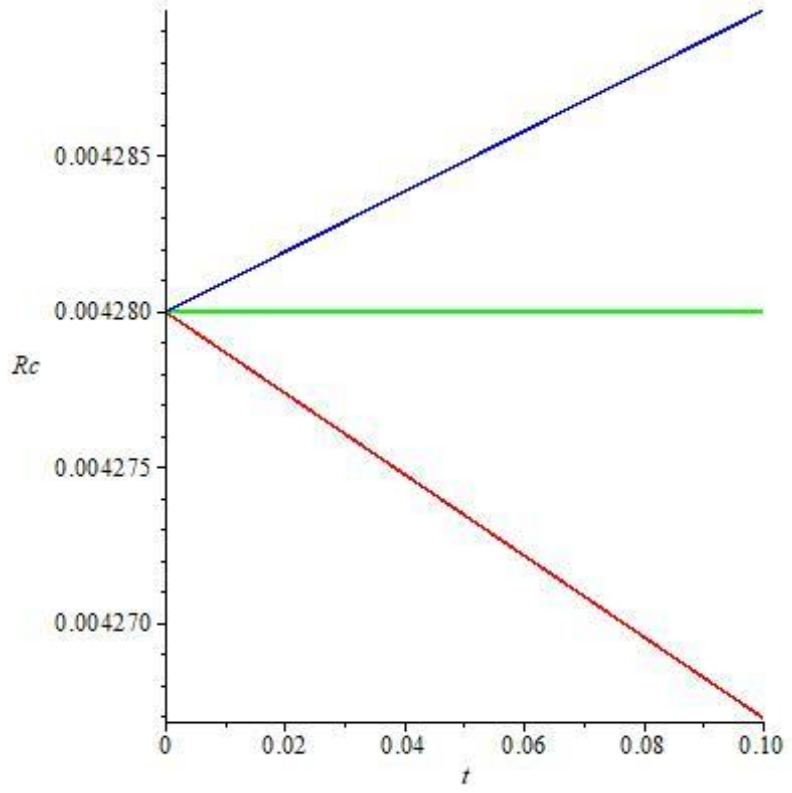

**Figure 4**.The inner meridian curve radius change for $v = 2\cdot 10^{-3}[m/s]$, $p_i = -4845\,[Pa]$ (red), $p_i = -4745\,[Pa]$ (green), and $p_i = -4645\,[Pa]$ (blue) , respectively.

For $v = 2\cdot 10^{-3}[m/s]$ and $\alpha_i(4.28\cdot 10^{-3}, p_i(t)) = \pi/2 - \alpha_g + 1\cdot 10^{-1}\sin(300\cdot t)$ the inner meridian curve radius evolution during the first $0.1[s]$ can be seen on Fig.5a.
On Fig.5.b the inner meridian curve radius evolution during the first $0.1[s]$ is presented for the considered instantaneous perturbations and for the above considered time dependent perturbation.

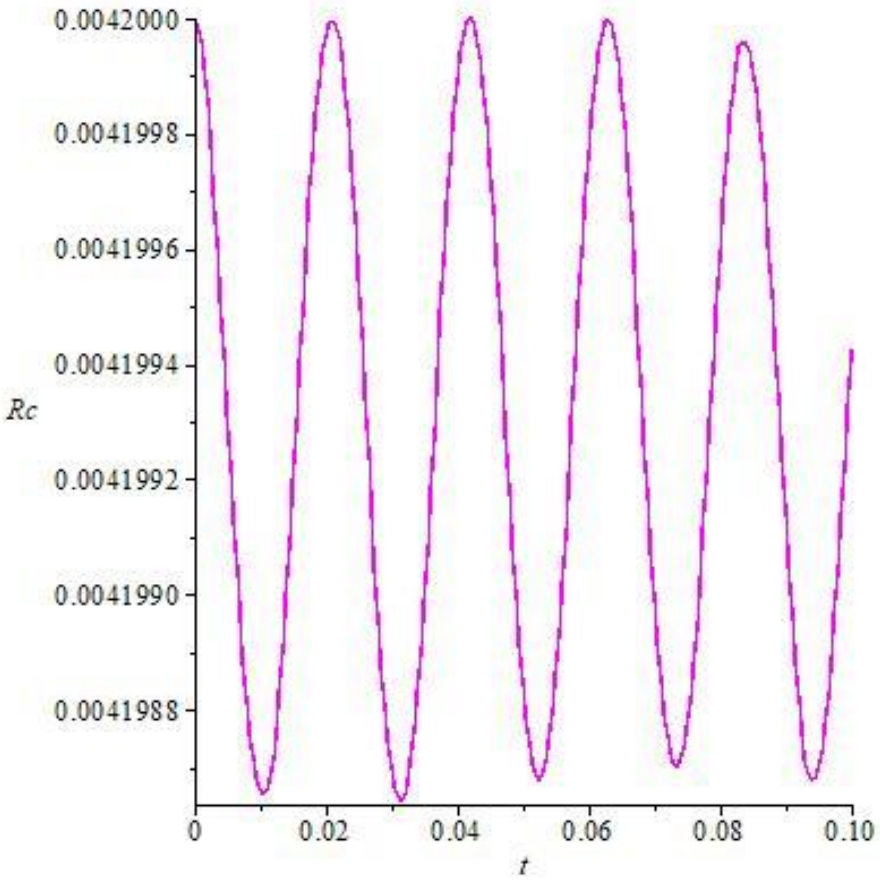

**Figure5a.** Inner meridian curve radius change for $v = 2\cdot 10^{-3}[m/s]$, $\alpha_i(4.28\cdot 10^{-3}, p_i(t)) = \pi/2 - \alpha_g + 1\cdot 10^{-1}\cdot\sin(300\cdot t)$

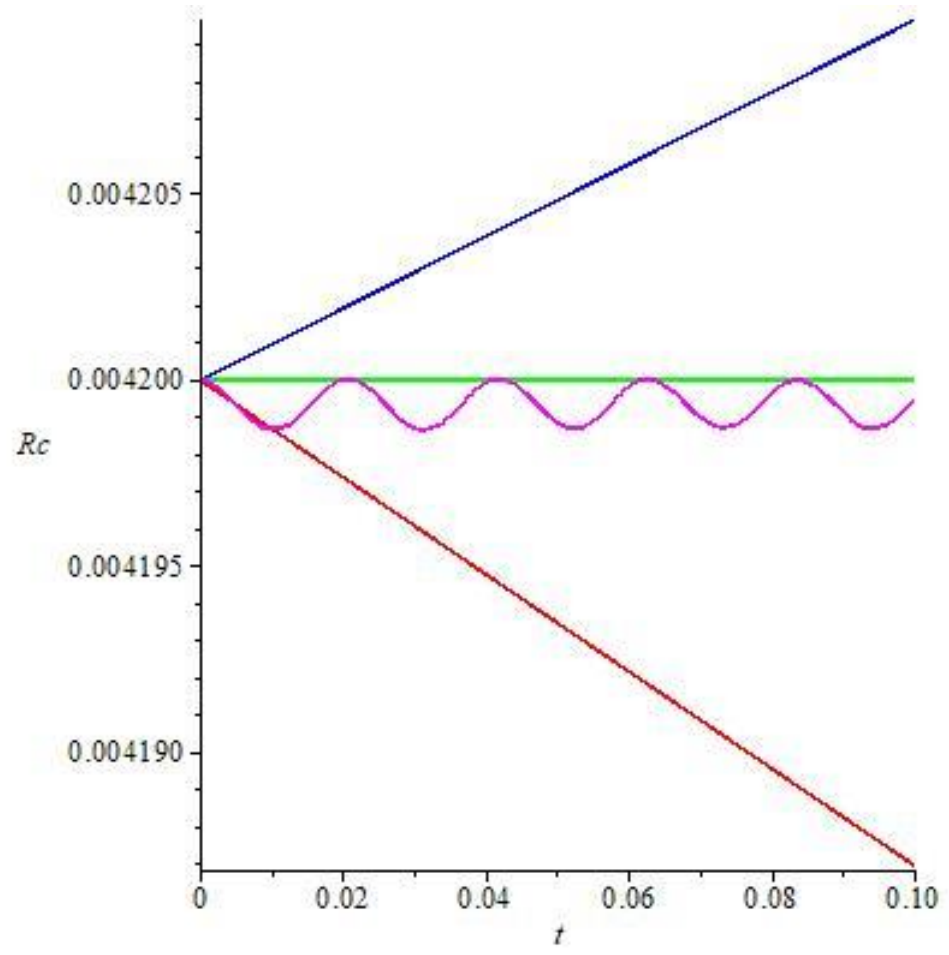

**Figure5b.** Inner meridian curve radius change for $v = 2\cdot10^{-3}[m/s]$, $p_i = -4745\,[Pa]$ (green), $p_i = -4845\,[Pa]$ (red), $p_i = -4645\,[Pa]$ (blue), $\alpha_i(4.28\cdot10^{-3}, p_i(t)) = \pi/2 - \alpha_g + 1\cdot10^{-1}\sin(300\cdot t)$ (magenta)

On Fig.5b. can be seen that:

- if in the first $0.1[s]$ inner pressure difference $p_i$ is constant equal to $-4745\,[Pa]$, then the inner meridian curve radius $R_{ci}$ is constant equal to $R_{ci} = 4.28\cdot10^{-3}[m]$ (color green).
- if at $t=0$,the inner pressure difference $p_i$ decreases instantaneously from $-4745\,[Pa]$ to $-4845\,[Pa]$, then the inner meridian curve radius decreases in $0.1[s]$ from $R_{c_i} = 4.28\cdot10^{-3}\,[m]$ to $R_c' = 4.26694\cdot10^{-3}[m]$ (color red)
- if at $t=0$,the inner pressure difference $p_i$ increases instantaneously from $-4745\,[Pa]$ to $-4645[Pa]$, then the inner meridian curve radius increases in $0.1[s]$ from $R_{ci} = 4.28\cdot10^{-3}[m]$ to $R_c'' = 4.28967\cdot10^{-3}[m]$ (color blue).
- if at $t=0$, the inner pressure difference becomes oscillatory, then the inner meridian curve radius oscillates too.

**v). The effect of the variation of pressure difference** $p_e$

Starting from the condition of growth angle constancy, according to [26]-Chapter 1, the equation of the outer radius change rate, due to the perturbation of the outer pressure difference $-4730[Pa]$ , is:

$$\begin{cases} dR_{ce}/dt = -v\tan(\alpha_e(4.72\cdot10^{-3}, p_e) - (\pi/2 - \alpha_g)) \\ R_{ce}(0) = 4.72\cdot10^{-3} \end{cases} \qquad (5.5)$$

Here $R_{ce}(t)$ is the inner radius at the moment of time $t$, $v = 2\cdot10^{-3}[m/s]$ is the pulling rate used in Eris experiment[1] , $\alpha_e(4.72\cdot10^{-3}, p_e)$ is the angle made by the line tangent to the perturbed inner meridian curve at the point of coordinates $(4.72\cdot10^{-3}, z_e(4.72\cdot10^{-3}, p_e))$ ,with the horizontal axis and $p_e$ is the value of the perturbed outer pressure difference.

If $p_e = -4730\,[Pa]$, then $\alpha_e(4.72\cdot10^{-3}, p_e) = \pi/2 - \alpha_g$ .

So, $\tan(\alpha_e(4.72\cdot10^{-3}, p_e) - (\pi/2 - \alpha_g)) = \tan(0) = 0$ and (5.1) becomes:

$$\begin{cases} dR_{ce}/dt = 0 \\ R_{ce}(0) = 4.72\cdot10^{-3} \end{cases} \qquad (5.6)$$

For $p_e = -4830\,[Pa]$ then $\alpha_e(4.72\cdot10^{-3}, p_e) = 1.443569\,[\mathrm{rad}]$.

So, $-v\cdot\tan(\alpha_e(4.72\cdot10^{-3},p_e)-(\pi/2-\alpha_g))=-v\cdot\tan(0.06477)$ and (5.1) becomes:

$$\begin{cases} dR_{ce}/dt=-v\tan(0.06477) \\ R_{ce}(0)=4.72\cdot10^{-3} \end{cases} \qquad (5.7)$$

For $p_e=-4630\,[Pa]$ we have $\alpha_e(4.72\cdot10^{-3},p_e)=1.33014[rad]$.

So, $-v\cdot\tan(\alpha_e(4.72\cdot10^{-3},p_e)-(\pi/2-\alpha_g))=-v\cdot\tan(-0.48656)$ and (5.1) becomes:

$$\begin{cases} dR_{ce}/dt=v\cdot\tan(0.04865) \\ R_{ce}(0)=4.72\cdot10^{-3} \end{cases} \qquad (5.8)$$

For $v=2\cdot10^{-3}[m/s]]$ the solutions of the initial value problems (5.6), (5.7) and (5.8) are represented on Fig.6.with green, red and blue, respectively.

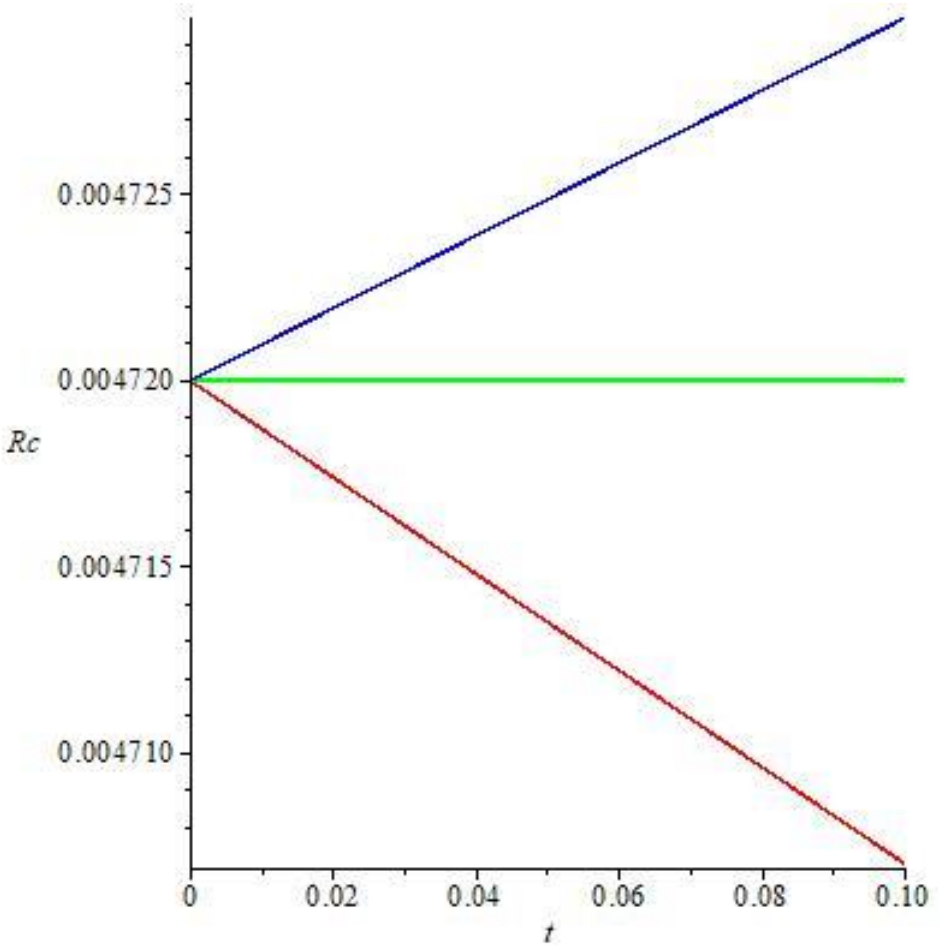

**Figure.6**.The outer meridian curve radius change for $v=2\cdot10^{-3}[m/s]$, $p_e=-4830\,[Pa]$ (red), $p_i=-4730\,[Pa]$ (green), and $p_i=-4630\,[Pa]$ (blue) , respectively.

For $v=2\cdot10^{-3}[m/s]$ and $\alpha_e(4.72\cdot10^{-3},\,p_e(t))=\pi/2-\alpha_g+1\cdot10^{-1}\sin(300\cdot t)$ the outer meridian curve radius evolution during the first $0.1[s]$ can be seen on Fig.7a.

On Fig.7.b the inner meridian curve radius evolution during the first $0.1[s]$ is presented for the considered instantaneous perturbations and for the above considered time dependent perturbation.

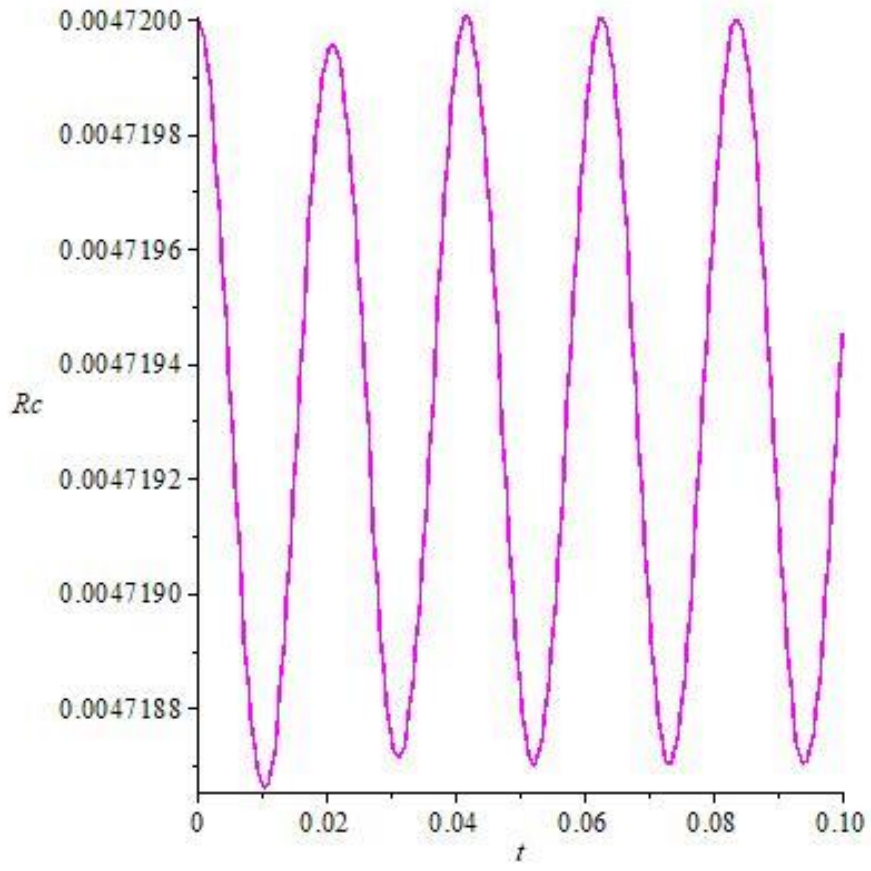

**Figure7a.**Inner meridian curve radius change for $v=2\cdot10^{-3}[m/s]$,

$\alpha_e(4.72\cdot10^{-3},p_e(t))=\pi/2-\alpha_g+1\cdot10^{-1}\cdot\sin(300\cdot t)$

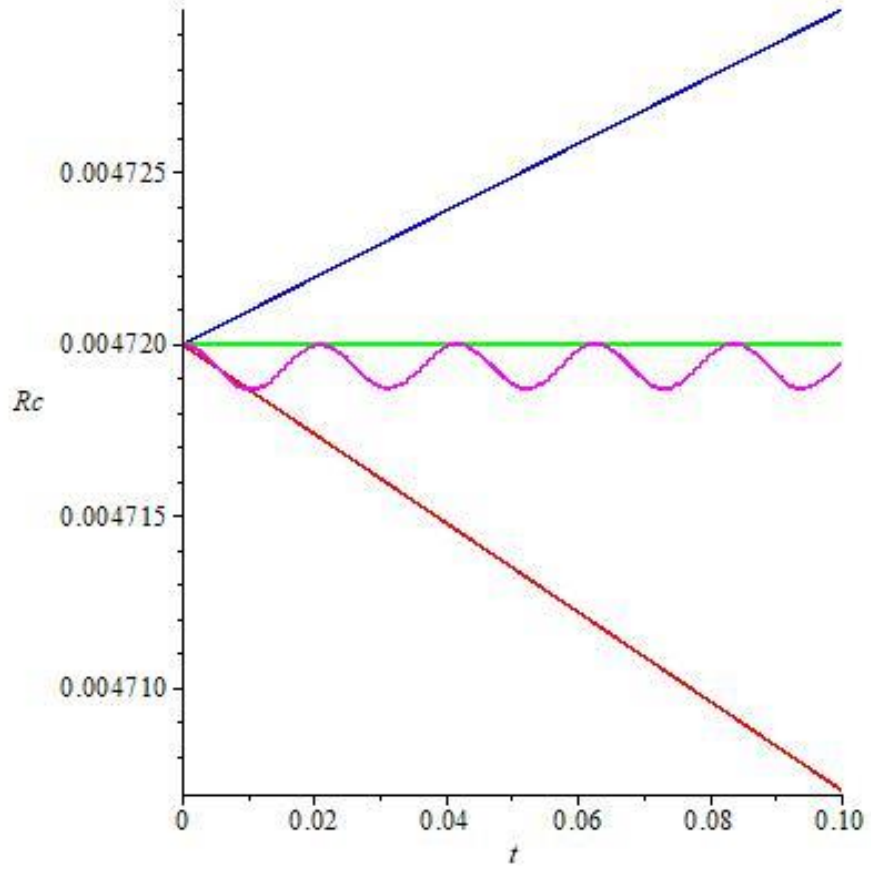

**Figure7b.** Inner meridian curve radius change for $v = 2 \cdot 10^{-3} [m/s]$ $p_e = -4730\,[Pa]$ (green), $p_e = -4830\,[Pa]$ (red), $p_e = -4630\,[Pa]$ (blue) , $\alpha_i(4.72 \cdot 10^{-3}, p_e(t)) = \pi/2 - \alpha_g + 1 \cdot 10^{-1} \sin(300 \cdot t)$ (magenta)

On Fig.7b. can be seen that:

- if in the first $0.1[s]$ outer pressure difference $p_e$ is constant equal to $-4730\,[Pa]$, then the outer meridian curve radius $R_{ei}$ is constant equal to $R_{ei} = 4.72 \cdot 10^{-3} [m]$ (color green).
- if at $t = 0$ the outer pressure difference $p_e$ decreases instantaneously from $-4730\,[Pa]$ to $-4830\,[Pa]$, then the outer meridian curve radius decreases in $0.1[s]$ from $R_{ce} = 4.72 \cdot 10^{-3}\,[m]$ to $R_c{}' = 4.70702 \cdot 10^{-3} [m]$ (color red)
- if at $t = 0$ the outer pressure difference $p_e$ increases instantaneously from $-4730\,[Pa]$ to $-4630[Pa]$, then the outer meridian curve radius increases in $0.1[s]$ from $R_{ce} = 4.72 \cdot 10^{-3} [m]$ to $R_c{}'' = 4.72973 \cdot 10^{-3} [m]$ (color blue).
- if at t=0 the inner pressure difference becomes oscillatory, then the inner meridian curve radius oscillates too.

## 6. CONCLUSIONS

1. In order to obtain an inner capillary free surface with convex meridian curve the inner pressure difference across the free surface has to be searched in the range defined by (3.2) and (3.3).In order to find the exact value of the inner pressure difference across the inner capillary surface the initial value problem (3.4) has to be solved for the values of the inner pressure difference in the range defined by (3.2),(3.3).
2. In order to obtain an outer capillary free surface with convex meridian curve the outer pressure difference across the surface has to be searched in the range defined by (3.6) and (3.7). In order to find the exact value of the outer pressure difference across the outer capillary surface the initial value problem (3.8) has to be solved for the values of the inner pressure difference in the range defined by (3.6), (3.7).
3. If inequality (4.1) holds, then the inner capillary free surface is static stable, i.e. such a capillary surface exists in real world. If inequality (4.2) holds, then the inner capillary free surface is static unstable, i.e. such a capillary surface exists only mathematically but doesn't exist in real world.
4. If inequality (4.3) holds, then the outer capillary free surface is static stable, i.e. such a capillary surface exists in real world. If inequality (4.4) holds, then the outer capillary free surface is static unstable, i.e. such a capillary surface exists only mathematically, but doesn't exist in real world.

5. In the case considered in numerical illustration the computed inner pressure difference is less than the outer pressure difference and the inner crystallization front height is more than the outer crystallization front height.

6. For the same numerical data in case of the E.F.G. growth method $p_i = -4743[Pa]$ and $p_e = -4490[Pa]$ instead of $p_i = -4745[Pa]$ and $p_e = -4730[Pa]$ in case of PD method.

**ACKNOWLEDGEMENT**

This research did not receive any specific grant from funding agencies in the public, commercial, or not-for-profit sectors.

## 7. APPENDIX

### 1. Proof of the Statement 1

Let $\alpha_c + \alpha_g < \pi/2$ , $R_{ci} = m \cdot R_{gi}$, $1 < m < \frac{R_{gi} + R_{ge}}{2 \cdot R_{gi}}$ and $z_i(r)$ defined for $r \in \lfloor R_{gi}, R_{ci} = m \cdot R_{gi} \rfloor$ which verifies (2.6), (2.8) - (2.10) and $z_i''(r) > 0$.The function defined as:

$$\alpha_i(r) = \arctan(z_i'(r)) \quad \text{for } r \in \lfloor R_{gi}, R_{ci} = m \cdot R_{gi} \rfloor$$

verifies

$$\alpha_i'(r) = -\frac{\rho \cdot g \cdot z_i(r) + p_i}{\gamma} \cdot \frac{1}{\cos(\alpha_i(r))} - \frac{1}{r} \cdot \tan(\alpha_i(r))$$

and the boundary conditions : $\alpha_i(R_{gi}) = \alpha_c$, $\alpha_i(R_{ci}) = \pi/2 - \alpha_g$ .
Hence, by the mean value theorem, there exists $r' \in \lfloor R_{gi}, R_{ci} \rfloor$ such that the following equality holds:

$$p_i = \gamma \cdot \frac{\alpha_c + \alpha_g - \pi/2}{R_{ci} - R_{gi}} \cdot \cos\alpha_i(r') - \rho \cdot g \cdot z_i(r') - \frac{\gamma}{r'} \cdot \sin(\alpha_i(r')$$

On the other hand, inequality $z_i''(r) > 0$ implies that the function $z'_i(r)$ is strictly increasing and by consequence, the function $\alpha_i(r)$ is strictly increasing. Therefore, $\alpha_c < \alpha_i(r') < \frac{\pi}{2} - \alpha_g$ .
Now, by taking into account that $z_i(r')$ >0 and $\gamma \cdot \frac{\alpha_c + \alpha_g - \pi/2}{R_{ci} - R_{gi}} < 0$ the following relations hold:

$$p_i = \gamma \cdot \frac{\alpha_c + \alpha_g - \pi/2}{R_{ci} - R_{gi}} \cdot \cos\alpha_i(r') - \rho \cdot g \cdot z_i(r') - \frac{\gamma}{r'} \cdot \sin(\alpha_i(r')) < \gamma \cdot \frac{\alpha_c + \alpha_g - \pi/2}{R_{ci} - R_{gi}} \cdot \cos\alpha_i(\frac{\pi}{2} - \alpha_g) -$$

$$\gamma \cdot \frac{1}{R_{ci}} \cdot \sin\alpha_c < \gamma \cdot \frac{\alpha_c + \alpha_g - \pi/2}{R_{gi}} \cdot \frac{1}{m-1} \sin\alpha(\alpha_g) - \frac{\gamma}{R_{gi}} \cdot \frac{1}{m} \cdot \sin(\alpha_c)$$

with $m = \frac{R_{ci}}{R_{gi}} > 1$

So, the right hand side of the inequality (3.1) is proven.
In order to obtain the left hand side of (3.1) remark first the inequality :

$$p_i = \gamma \cdot \frac{\alpha_c + \alpha_g - \pi/2}{R_{ci} - R_{gi}} \cdot \cos\alpha_i(r') - \rho \cdot g \cdot z_i(r') - \frac{\gamma}{r'} \cdot \sin(\alpha_i(r')) > \gamma \cdot \frac{\alpha_c + \alpha_g - \pi/2}{R_{ci} - R_{gi}} \cdot \cos\alpha_c -$$

$$\rho \cdot g \cdot z_i(R_{ci}) - \frac{\gamma}{R_{gi}} \cdot \cos(\alpha_g).$$

The term $-\rho \cdot g \cdot z_i(R_{ci})$ can be evaluated applying the mean value theorem for the function $z_i(r)$ on $\lfloor R_{gi}, R_{ci} \rfloor$ . It follows that there exists $r'' \in \lfloor R_{gi}, R_{ci} \rfloor$ such that $z_i(R_{ci}) - z_i(R_{gi}) = (R_{ci} - R_{gi}) z_i'(r'') > (R_{ci} - R_{gi}) z_i'(R_{gi}) = (R_{ci} - R_{gi}) \tan(\alpha_c)$ with $z_i(R_{gi}) = 0$ .
Consequently the following inequality holds:

$$-\rho \cdot g \cdot z_i(R_{ci}) < -\rho \cdot g \cdot (R_{ci} - R_{gi}) \tan(\alpha_c) .$$

So, for $p_i$ the following relations hold:

$$p_i > \gamma \cdot \frac{\alpha_c + \alpha_g - \pi/2}{R_{ci} - R_{gi}} \cdot \cos\alpha_c - \rho \cdot g \cdot (R_{ci} - R_{gi}) \cdot \tan(\alpha_c) - \frac{\gamma}{R_{gi}} \cdot \cos(\alpha_g).$$

$$> \gamma \cdot \frac{\alpha_c + \alpha_g - \pi/2}{R_{gi}} \cdot \frac{1}{m-1} \cdot \cos\alpha_c - \rho \cdot g \cdot R_{gi} \cdot (m-1) \cdot \tan(\alpha_c) - \frac{\gamma}{R_{gi}} \cdot \frac{1}{m} \cdot \cos(\alpha_g)$$

This means that the left hand side of (3.1) is proven.

**2. Proof of the Statement 2**

Let $\alpha_c + \alpha_g < \pi/2$ , $R_{ce} = \frac{1}{n} \cdot R_{ge}$, $1 < n < \frac{2 \cdot R_{ge}}{R_{gi} + R_{ge}}$ and $z_e(r)$ defined for $r \in \left[R_{ce} = \frac{1}{n} R_{ge}, R_{ge}\right]$ which verifies (2.7), (2.12) - (2.14) and $z_e''(r) > 0$ The function defined as:

$$\alpha_e(r) = -\arctan(z_e'(r)) \quad \text{for } r \in \left[R_{ce} = \frac{1}{n} \cdot R_{ge}, R_{ge}\right]$$

verifies

$$\alpha_e'(r) = \frac{\rho \cdot g \cdot z_e(r) + p_e}{\gamma} \cdot \frac{1}{\cos(\alpha_e(r))} + \frac{1}{r} \cdot \tan(\alpha_e(r))$$

and the boundary conditions : $\alpha_e(R_{ge}) = \alpha_c$ , $\alpha_e(R_{ce}) = \pi/2 - \alpha_g$ .

Hence, by the mean value theorem, there exists $r' \in \lfloor R_{ce}, R_{ge} \rfloor$ such that the following equality holds:

$$p_e = \gamma \cdot \frac{\alpha_c + \alpha_g - \pi/2}{R_{ge} - R_{ce}} \cdot \cos\alpha_e(r') - \rho \cdot g \cdot z_e(r') - \frac{\gamma}{r'} \cdot \sin(\alpha_e(r')).$$

On the other hand, inequality $z_e''(r) > 0$ implies that the function $z'_e(r)$ is strictly increasing and by consequence the function $\alpha_e(r)$ is strictly decreasing. Therefore, $\alpha_c < \alpha_e(r') < \frac{\pi}{2} - \alpha_g$ .

Now, by taking into account that $z_e(r') < 0$ and $\gamma \cdot \frac{\alpha_c + \alpha_g - \pi/2}{R_{ge} - R_{ce}} < 0$ the following relations hold:

$$p_e = \gamma \cdot \frac{\alpha_c + \alpha_g - \pi/2}{R_{ge} - R_{ce}} \cdot \cos\alpha_e(r') - \rho \cdot g \cdot z_e(r') - \frac{\gamma}{r'} \cdot \sin(\alpha_e(r')) < \gamma \cdot \frac{\alpha_c + \alpha_g - \pi/2}{R_{ge} - R_{ce}} \cdot \cos\alpha_e(\frac{\pi}{2} - \alpha_g) -$$

$$\gamma \cdot \frac{1}{R_{ge}} \cdot \sin\alpha_c < \gamma \cdot \frac{\alpha_c + \alpha_g - \pi/2}{R_{ge}} \cdot \frac{n}{n-1} \sin(\alpha_g) - \frac{\gamma}{R_{ge}} \cdot \sin(\alpha_c)$$

with $n = \frac{R_{ge}}{R_{ce}} > 1$

So, the right hand side of the inequality (3.2) is proven.

In order to obtain the left hand side of (3.1) remark first the inequality :

$$p_e = \gamma \cdot \frac{\alpha_c + \alpha_g - \pi/2}{R_{ge} - R_{ce}} \cdot \cos\alpha_e(r') - \rho \cdot g \cdot z_e(r') - \frac{\gamma}{r'} \cdot \sin(\alpha_e(r')) > \gamma \cdot \frac{\alpha_c + \alpha_g - \pi/2}{R_{ge} - R_{ce}} \cdot \cos\alpha_c -$$

$$\rho \cdot g \cdot z_e(R_{ce}) - \frac{\gamma}{R_{ce}} \cdot \cos(\alpha_g).$$

The term $-\rho \cdot g \cdot z_e(R_{ce})$ can be evaluated applying the mean value theorem for the function $z_e(r)$ on $\lfloor R_{ce}, R_{ge} \rfloor$ . It follows that there exists $r'' \in \lfloor R_{ce}, R_{ge} \rfloor$ such that $z_e(R_{ge}) - z_e(R_{ce}) = (R_{ge} - R_{ce}) z_e'(r'') > (R_{ge} - R_{ce}) z_e'(R_{ge}) = -(R_{ge} - R_{ce}) \cdot \tan(\frac{\pi}{2} - \alpha_g)$ with $z(R_{ge}) = 0$ .

So, for $p_e$ the following relations hold:

$$p_e > \gamma \cdot \frac{\alpha_c + \alpha_g - \pi/2}{R_{ge} - R_{ce}} \cdot \cos\alpha_c - \rho \cdot g \cdot (R_{ge} - R_{ce}) \cdot \tan(\alpha_c) - \frac{\gamma}{R_{ce}} \cdot \cos(\alpha_g)$$

$$> \gamma \cdot \frac{\alpha_c + \alpha_g - \pi/2}{R_{ge}} \cdot \frac{n}{n-1} \cdot \cos\alpha_c - \rho \cdot g \cdot R_{ge} \cdot \frac{n-1}{n} \cdot \tan(\frac{\pi}{2} - \alpha_g) - \frac{\gamma}{R_{ge}} \cdot n \cdot \cos(\alpha_g)$$

This means that the left hand side of (3.2) is proven.

**3. Proof of the Statement 3**

Since (2.6) is the Euler equation [9] for the free energy functional (2.11), in this case it is sufficient to investigate the Legendre and Jacobi conditions [9] .
Consider for that the function

$$F(z_i, z_i', r) = \left\{ \gamma \cdot \left[1 + (z_i')^2\right]^{1/2} - \frac{1}{2} \cdot \rho \cdot g \cdot z_i^2 - p_i \cdot z_i \right\} \cdot r$$

and remark that the Legendre condition $\frac{\partial^2 F}{\partial^2 z_i'} > 0$ reduces to the inequality:

$$r \cdot \gamma \left[1 + (z_i')^2\right]^{-3/2} > 0$$

which is satisfied .
The Jacobi equation

$$\left[\frac{\partial^2 F}{\partial z_i^2} - \frac{d}{dr}\left(\frac{\partial^2 F}{\partial z_i \partial z_i'}\right)\right] \cdot \eta - \frac{d}{dr}\left[\frac{\partial^2 F}{\partial z'^2} \cdot \eta'\right] = 0$$

in this case become

$$\frac{d}{dr}\left(\frac{r \cdot \gamma}{[1 + (z_i')^2]^{3/2}} \cdot \eta'\right) + \rho \cdot g \cdot r \cdot \eta = 0$$

i).For obtaining the stability result remark that if $z_i(r)$ satisfies conditions (2,6),(2.8),(2.9),(2.10) and is convex ($z''_i(r) > 0$), then for the coefficients of the Jacobi equation the following inequalities hold:

$$\frac{r \cdot \gamma}{[1 + (z_i')^2]^{3/2}} \geq R_{gi} \cdot \gamma \cdot \sin^3 \alpha_g \qquad\qquad \rho \cdot g \cdot r \leq \rho \cdot g \cdot m \cdot R_{gi}$$

Therefore the equation

$$\frac{d}{dr}\left(R_{gi} \cdot \gamma \cdot \sin^3 \alpha_g \cdot \varsigma'\right) + \rho \cdot g \cdot m \cdot R_{gi} \cdot \varsigma = 0$$

is a „Sturm type upper bound" [10] for the Jacobi equation . An arbitrary solution of the above „Sturm type upper bound equation" is given by

$$\varsigma(r) = A \cdot \sin(\omega \cdot r + \varphi)$$

Here $A$ and $\varphi$ are arbitrary real constants and $\omega^2 = \frac{\rho \cdot g \cdot m}{\gamma \cdot \sin^3 \alpha_g}$.The half period of any non zero solution $\varsigma(r)$ is $\frac{\pi}{\omega} = \pi \cdot \frac{\gamma^{1/2} \cdot \sin^{3/2} \alpha_g}{\rho^{1/2} \cdot g^{1/2} \cdot m^{1/2}}$. If the half period is greater than $R_{ci} - R_{gi}$, then any non zero solution $\varsigma(r)$ vanishes at most once on the interval $[R_{gi}, R_{ci}]$. In other words, if the following inequalities hold

$$\pi \cdot \frac{\gamma^{1/2} \cdot \sin^{3/2} \alpha_g}{\rho^{1/2} \cdot g^{1/2} \cdot m^{1/2}} > R_{gi} \cdot (m-1) \quad \text{or} \quad \frac{m-1}{m^{1/2}} < \pi \cdot \frac{1}{R_{gi}} \cdot \frac{\gamma^{1/2} \cdot \sin^{3/2} \alpha_g}{\rho^{1/2} \cdot g^{1/2}},$$

then any non zero solution $\varsigma(r)$ vanishes at most once on the interval $[R_{gi}, R_{ci}]$ . Hence, according to [10] Chapter11, the solution $\eta(r)$ of Jacobi equation which satisfies $\eta(R_{gi}) = 0$ and $\eta'(R_{gi}) = 1$ has only one zero on the interval $[R_{gi}, R_{ci}]$. This means that the Jacobi condition for weak minimum is satisfied [9].

ii).For obtaining the instability result remark now that for the coefficients of the Jacobi equation the following inequalities hold:

$$\frac{r \cdot \gamma}{[1+(z')^2]^{3/2}} \leq m \cdot R_{ge} \cdot \gamma \cdot \cos^3 \alpha_c \qquad \text{and} \qquad \rho \cdot g \cdot r \geq \rho \cdot g \cdot R_{ge}$$

Therefore the equation

$$\frac{d}{dr}(m \cdot R_{ge} \cdot \gamma \cdot \cos^3 \alpha_c \cdot \xi') + \rho \cdot g \cdot R_{ge} \cdot \xi = 0$$

is a „Sturm type lower bound equation" [10] for the Jacobi equation . An arbitrary solution of the above „Sturm type lower bound equation" is given by

$$\xi(r) = A \cdot \sin(\omega \cdot r + \varphi)$$

Here $A$ and $\varphi$ are arbitrary real constants and $\omega^2 = \dfrac{\rho \cdot g}{m \cdot \gamma \cdot \cos^3 \alpha_c}$. The period of any non zero solution $\xi(r)$ is $\dfrac{2 \cdot \pi}{\omega} = 2 \cdot \pi \cdot \dfrac{\gamma^{1/2} \cdot m^{1/2} \cos^{3/2} \alpha_c}{\rho^{1/2} \cdot g^{1/2}}$. If the period is less than $R_{ci} - R_{gi}$, then any non zero solution $\xi(r)$ vanishes at least twice on the interval $[R_{gi}, R_{ci}]$.I n other words, if the following inequalities hold :

$$2 \cdot \pi \cdot \frac{\gamma^{1/2} \cdot m^{1/2} \cos^{3/2} \alpha_c}{\rho^{1/2} \cdot g^{1/2}} < R_{gi} \cdot (m-1) \quad \text{or} \quad \frac{m-1}{m^{3/2}} > 2 \cdot \pi \cdot \frac{1}{R_{gi}} \cdot \frac{\gamma^{1/2} \cdot \cos^{3/2} \alpha_c}{\rho^{1/2} \cdot g^{1/2}},$$

then any non zero solution $\xi(r)$ vanishes at least twice on the interval $[R_{gi}, R_{ci}]$. Hence, according to [10], the solution $\eta(r)$ of Jacobi equation which satisfies $\eta(R_{gi}) = 0$ and $\eta'(R_{gi}^{'}) = 1$ has at least two zero on the interval $[R_{gi}, R_{ci}]$.This means that the Jacobi condition for weak minimum is not satisfied [9].

**4. Proof of the Statement 4**

Since (2.7) is the Euler equation [9] for the free energy functional (2.15), in this case it is sufficient to investigate the Legendre and Jacobi conditions [9].

Consider for that the function

$$F(z_e, z_e', r) = \left\{ \gamma \cdot \left[1 + (z_e')^2\right]^{1/2} - \frac{1}{2} \cdot \rho \cdot g \cdot z_e^2 - p_e \cdot z_e \right\} \cdot r$$

and remark that the Legendre condition $\dfrac{\partial^2 F}{\partial^2 z_e'} > 0$ reduces to the inequality:

$$r \cdot \gamma \left[1 + (z_e')^2\right]^{-3/2} > 0$$

which is satisfied .

The Jacobi equation

$$\left[\frac{\partial^2 F}{\partial z^2} - \frac{d}{dr}\left(\frac{\partial^2 F}{\partial z_e \partial z_e'}\right)\right] \cdot \eta - \frac{d}{dr}\left[\frac{\partial^2 F}{\partial z_e'^2} \cdot \eta\right]' = 0$$

in this case becomes

$$\frac{d}{dr}\left(\frac{r \cdot \gamma}{[1 + (z_e')^2]^{3/2}} \cdot \eta'\right) + \rho \cdot g \cdot r \cdot \eta = 0$$

i).For obtaining the stability result remark that using (2.7),(2.12)-(2.14) and $z_e''(r) > 0$ for the coefficients of the Jacobi equation the following inequalities can be obtained:

$$\frac{r \cdot \gamma}{[1+(z_e')^2]^{3/2}} \geq \frac{R_{ge}}{n} \cdot \gamma \cdot \sin^3 \alpha_g \qquad \rho \cdot g \cdot r \leq \rho \cdot g \cdot R_{ge}$$

Therefore the equation

$$\frac{d}{dr}\left(\frac{R_{ge}}{n} \cdot \gamma \cdot \sin^3 \alpha_g \cdot \varsigma'\right) + \rho \cdot g \cdot R_{ge} \cdot \varsigma = 0$$

is a „Sturm type upper bound" [10] for the Jacobi equation . An arbitrary solution of the above „Sturm type upper bound equation" is given by

$$\varsigma(r) = A \cdot \sin(\omega \cdot r + \varphi)$$

Here $A$ and $\varphi$ are arbitrary real constants and $\omega^2 = \frac{\rho \cdot g \cdot n}{\gamma \cdot \sin^3 \alpha_g}$. The half period of any non zero solution $\varsigma(r)$ is $\frac{\pi}{\omega} = \pi \cdot \frac{\gamma^{1/2} \cdot \sin^{3/2} \alpha_g}{\rho^{1/2} \cdot g^{1/2} \cdot n^{1/2}}$. If the half period is greater than $R_{ge} - R_{ce}$, then any non zero solution $\varsigma(r)$ vanishes at most once on the interval $[R_{ce}, R_{ge}]$. In other words, if the following inequalities hold

$$\pi \cdot \frac{\gamma^{1/2} \cdot \sin^{3/2} \alpha_g}{\rho^{1/2} \cdot g^{1/2} \cdot n^{1/2}} > R_{ge} \cdot (1 - \frac{1}{n}) \quad \text{or} \quad \frac{n-1}{n^{1/2}} < \pi \cdot \frac{1}{R_{ge}} \cdot \frac{\gamma^{1/2} \cdot \sin^{3/2} \alpha_g}{\rho^{1/2} \cdot g^{1/2}},$$

then any non zero solution $\varsigma(r)$ vanishes at most once on the interval $[R_{ce}, R_{ge}]$ . Hence, according to [10], the solution $\eta(r)$ of Jacobi equation which satisfies $\eta(R_{gi}) = 0$ and $\eta'(R_{gi}) = 1$ has only one zero on the interval $[R_{ce}, R_{ge}]$.This means that the Jacobi condition for weak minimum is satisfied [9].

ii).For obtaining the instability result remark now that for the coefficients of the Jacobi equation the following inequalities hold:

$$\frac{r \cdot \gamma}{[1+(z_e')^2]^{3/2}} \leq R_{ge} \cdot \gamma \cdot \cos^3 \alpha_c \quad \text{and} \quad \rho \cdot g \cdot r \geq \rho \cdot g \cdot \frac{R_{ge}}{n}$$

Therefore the equation

$$\frac{d}{dr}(R_{ge} \cdot \gamma \cdot \cos^3 \alpha_c \cdot \xi') + \rho \cdot g \cdot \frac{R_{ge}}{n} \cdot \xi = 0$$

is a „Sturm type lower bound equation" [10] for the Jacobi equation. An arbitrary solution of the above „Sturm type lower bound equation" is given by

$$\xi(r) = A \cdot \sin(\omega \cdot r + \varphi)$$

Here $A$ and $\varphi$ are arbitrary real constants and $\omega^2 = \frac{\rho \cdot g}{n \cdot \gamma \cdot \cos^3 \alpha_c}$. The period of any non zero solution $\xi(r)$ is $\frac{2 \cdot \pi}{\omega} = 2 \cdot \pi \cdot \frac{\gamma^{1/2} \cdot n^{1/2} \cos^{3/2} \alpha_c}{\rho^{1/2} \cdot g^{1/2}}$. If the period is less than $R_{ge} - R_{ce}$, then any non zero solution $\xi(r)$ vanishes at least twice on the interval $[R_{ce}, R_{ge}]$. In other words, if the following inequalities hold :

$$2 \cdot \pi \cdot \frac{\gamma^{1/2} \cdot n^{1/2} \cos^{3/2} \alpha_c}{\rho^{1/2} \cdot g^{1/2}} < R_{ge} \cdot (1 - \frac{1}{n}) \quad \text{or} \quad \frac{n-1}{n^{3/2}} > 2 \cdot \pi \cdot \frac{1}{R_{ge}} \cdot \frac{\gamma^{1/2} \cdot \cos^{3/2} \alpha_c}{\rho^{1/2} \cdot g^{1/2}}$$

then any non zero solution $\xi(r)$ vanishes at least twice on the interval $[R_{ce}, R_{ge}]$. Hence, according to [10], the solution $\eta(r)$ of Jacobi equation which satisfies $\eta(R_{gi}) = 0$ and $\eta(R'_{gi}) = 1$ has at least two zero on the interval $[R_{ce}, R_{ge}]$.This means that the Jacobi condition for weak minimum is not satisfied [9].